\documentclass[12pt, nonatbib]{elsarticle}

\usepackage{amssymb}
\usepackage{amsmath}
\usepackage{caption}
\usepackage[hidelinks]{hyperref}
\usepackage{subcaption}
\usepackage{soul}
\usepackage{color}
\usepackage{setspace}
\usepackage{booktabs}
\renewcommand\hl[1]{#1}
\makeatletter 
\let\c@author\relax
\makeatother
\usepackage[backend=biber,style=chem-angew,hyperref=true, articletitle=true, doi=true]{biblatex}
\journal{Journal of Magnetic Resonance}

\begin{document}

\begin{frontmatter}

\title{Deep Learning as a Method for Inversion of NMR Signals}

\author[1,2,3]{Julian B. B. Beckmann \corref{cor1}}
\ead{jbbeckmann@mgh.harvard.edu}
\author[1]{Mick D. Mantle}
\author[1]{Andrew J. Sederman}
\author[1]{Lynn F. Gladden} 

\affiliation[1]{organization={University of Cambridge, Department of Chemical Engineering and Biotechnology},
            addressline={Philippa Fawcett Drive}, 
            city={Cambridge},
            postcode={CB3 0AS}, 
            country={United Kingdom}}
\affiliation[2]{organization={Harvard Medical School},
            addressline={25 Shattuck Street}, 
            city={Boston},
            postcode={MA 02115}, 
            country={United States}}
\affiliation[3]{organization={Martinos Centre for Biomedical Imaging, Massachusetts General Hospital},
            addressline={149 13th Street}, 
            city={Charlestown},
            postcode={MA 02129}, 
            country={United States}}
            
\cortext[cor1]{Corresponding author}    

\begin{abstract}
The concept of deep learning is employed for the inversion of NMR signals and it is shown that NMR signal inversion can be considered as an image-to-image regression problem, which can be treated with a convolutional neural net. It is further outlined, that inversion through deep learning provides a clear efficiency and usability advantage compared to regularization techniques such as Tikhonov and modified total generalized variation (MTGV), because no hyperparemeter selection prior to reconstruction is necessary. The inversion network is applied to simulated NMR signals and the results compared with Tikhonov- and MTGV-regularization. The comparison shows that inversion via deep learning is significantly faster than the latter regularization methods and also outperforms both regularization techniques in nearly all instances.
\end{abstract}


\end{frontmatter}

\section{Introduction}
\label{sec:Int_ML}

In recent years, deep learning revolutionized many fields such as computer vision,\supercite{Krizhevsky2017ImageNetNetworks, Christian2015GoingConvolutions, He2015DeepRecognition} natural language processing\supercite{Mikolov2010RecurrentModel, Mikolov2011ExtensionsModel} or search recommendation prediction\supercite{Sen2016DeepRecommendations} and reached or even exceeded human performance for certain tasks.\supercite{He2015DelvingClassification} The concept of deep neural networks is several decades old, but recent progress in parallel computing\supercite{Krizhevsky2014OneNetworks} as well as the availability of vast data sets \supercite{Rajkomar2018ScalableRecords, Ray2015Privacy-PreservingLearning} made it possible to create deep enough models,\supercite{Lin1996LearningNetworks} which are able to make accurate predictions from complicated data such as traffic images\supercite{Isola2017Image-to-ImageNetworks} or voice recordings.\supercite{Mikolov2010RecurrentModel, Mikolov2011ExtensionsModel} As a result of the high commercial interest in fields such as self driving or machine translation, deep learning research was focused on computer vision and natural language processing, but recently research questions fundamental to the natural sciences\supercite{Raissi2019Physics-informedEquations, Qiao2020OrbNet:Features, Maziar2018DeepEquations, Qu2020AcceleratedLearning} and related areas such as engineering\supercite{Jha2018ElemNet:Composition} or medicine\supercite{Ribli2018DetectingLearning} gained interest from the deep learning community and vice versa. For instance, deep learning was very successfully employed for the prediction of protein folding\supercite{Jumper2021HighlyAlphaFold, Wei2019ProteinAlphaFold} or the identification of new medical compounds,\supercite{Stokes2020ADiscovery, Shen2021Out-of-the-boxRepresentations} but overall the application of machine learning and similar techniques to problems in natural science is still developing. This leads to the main purpose of this paper, which is the employment of deep learning as an inversion method for NMR signals. For a vast variety of NMR experiments, signal processing can be considered as an inverse problem.\supercite{PaulCallaghan2011TranslationalResonance, Song2002T1T2Inversion, Venkataramanan2002SolvingDimensions} In more detail, the signal acquired in multi-dimensional correlation experiments such as exchange-experiments, but also in simple one-dimensional diffusion measurements are sampled as a Laplace transform of the underlying physical distribution, which is usually the purpose of the experiment to acquire.\supercite{PaulCallaghan2011TranslationalResonance, Guo2019NuclearFunctions} Due to the ill-conditioned nature of the Laplace transform, no direct inverse transform \hl{as is} the case for the Fourier transform does exist and numerical inversion methods have to be employed to reconstruct the sought-after distribution from the acquired signal. \supercite{PaulCallaghan2011TranslationalResonance, Song2002T1T2Inversion, Venkataramanan2002SolvingDimensions, Guo2019NuclearFunctions} Optimization of these reconstruction problems has received a lot of scientific interest across a multitude of disciplines ranging from pure statistics to applied image processing.\supercite{Lawson1995SolvingProblems, Hastie2009TheLearning, Gareth2021AnLearning} In the case of NMR signal processing, the most prominent method to tackle the ill-conditioned nature of the Laplace transform is Tikhonov regularization.\supercite{Tikhonov1978SolutionsProblems.,Fuhry2012AMethod,Golub1999TikhonovSquares} The advantage of the Tikhonov method is its stability regarding noise, but simultaneously it suffers from poor resolution regarding closely aligned components and "over-smoothing" of sparse features.\supercite{Mitchell2012NumericalDimensions} To overcome these issues, multiple other regularisation techniques such as $\mathrm{L_1}$-regularization have been proposed and applied in the past.\supercite{Reci2017ObtainingProblems} The benefits of $\mathrm{L_1}$-regularization are its ability to separate closely aligned components as well as the correct reconstruction of sparse features, but at the same time $\mathrm{L_1}$-regularization shows poor performance regarding smooth features and a higher proneness of imposing reconstruction artefacts.\supercite{Reci2017ObtainingProblems} An alternative to $\mathrm{L_1}$- and Tikhonov regularization is modified total generalized variation regularization (MTGV).\supercite{Reci2017RetainingExperiments., Beckmann_MTGV} MTGV employs two separate penalty terms in its cost function. The first one ensuring sparsity and the second term enforcing smoothness on the reconstructed distribution. This means that sparse as well as smooth features can be reconstructed and the ratio between sparsity and smoothness can be controlled via an additional hyperparameter.\supercite{Reci2017RetainingExperiments., Beckmann_MTGV} The drawback of all three methods is that prior to the reconstruction the regularization hyperparameters have to be estimated. To date, there is no general consensus on the optimal method for selecting suitable hyperparameters and a significant number of methods for hyperparameter selection including generalized cross-validation\supercite{Golub1979GeneralizedParameter, Wen2018UsingProblems, Beckmann_MTGV} as well as the Butler-Reeds-Dawson method\supercite{Butler1981EstimatingSmoothing, Beckmann_MTGV} can be found in the literature,\supercite{Orr1996IntroductionNetworks} but in general the following two issues regarding hyperparameter selection continue to be persistent. Firstly, the vast majority of selection methods are based on the convergence of its employed metric and consequently, the final hyperparameter selection depends on the choice of a convergence or stopping criterion. Those criteria can be chosen empirically, but eventually every selection can be considered as a trade-off between accuracy and convergence speed.\supercite{Hastie2009TheLearning, Gareth2021AnLearning} This means that multiple selections of hyperparameters are valid and \hl{hence} some physical background knowledge is still needed to decide which value provides the best reconstruction from a physical perspective. The second disadvantage stemming from hyperparameter selection is that independent of the method employed a significant number of unused reconstructions has to be generated decreasing the final efficiency of the inversion technique considerably.\supercite{Mitchell2012NumericalDimensions} Both of these issues can be circumvented if inversion is conducted via deep learning instead \hl{of regularization} techniques. In more detail, for the inversion of the NMR signal with an already trained deep neural network no estimation of hyperparameters is necessary, which means the signal data can be passed right away to the network, which finally reconstructs the sought-after distribution.\supercite{Hastie2009TheLearning, Gareth2021AnLearning} This also prevents the reconstruction of unnecessary distributions during a hyperparameter search, which significantly increases the efficiency of the inversion method. Additionally, the lacking necessity for a hyperparameter selection method simplifies the overall inversion procedure, which clearly improves its usability as a routine inversion method. 
\section{Theoretical Background}
\label{sec:Theo_ML}

The NMR signal in a $T_1$-$T_2$-experiment is given as the Laplace transform of the sought-after distribution and can be described by the following equation:\supercite{Mitchell2012NumericalDimensions, Reci2017ObtainingProblems, Reci2017RetainingExperiments., PaulCallaghan2011TranslationalResonance}
\begin{equation} \label{eq:Sig_Math}
    \begin{split}
        S(t_1, t_2) = & \int_{0}^{\infty} \int_{0}^{\infty} F(T_1, T_2) K_1(t_1, T_1) K_2(t_2, T_2)... \; dT_1 \, dT_2... \\
        & + E(t_1, t_2),
        \end{split}
\end{equation}
where $F$ is the $T_1$-$T_2$-distribution, $K_{1,2}$ the kernel functions given by some sort of exponential decays with their exact definitions depending on the experiment, $E$ the inherent noise in the signal and $t_{1,2}$ are the time delays used to encode for $T_{1,2}$ in the NMR pulse sequence. Equation \ref{eq:Sig_Math} can be brought into discrete form given by the following expression:
\begin{equation} \label{eq:SigDis}
    \underline{\mathbf{S}} = \underline{\mathbf{K}}\,\underline{\mathbf{F}} + \underline{\mathbf{E}},
\end{equation}
Hence, the inversion problem which has to be solved is to estimate the distribution~$\underline{\mathbf{F}}$, while the kernel~$\underline{\mathbf{K}}$ and the NMR signal~$\underline{\mathbf{S}}$ are known. In the case a regularization approach is employed this leads to the following minimisation problem:\supercite{Tikhonov1978SolutionsProblems.,Fuhry2012AMethod, Golub1999TikhonovSquares}
\begin{equation} \label{eq:Tik_ML}
    \underline{\mathbf{F}} = \arg \; \min {}_{\underline{\mathbf{F}} \, \geq \, 0} \left(\frac{\alpha}{2} ||\underline{\mathbf{K}} \, \underline{\mathbf{F}} - \underline{\mathbf{S}}||_2^2 + P\left(\underline{\mathbf{F}}\right) \right),
\end{equation}
where $P\left(\underline{\mathbf{F}}\right)$ is some sort of regularization penalty and  $\alpha$ is a hyperparameter controlling the ratio between the penalty term and fidelity of the reconstructed distribution. Common choices for $P\left(\underline{\mathbf{F}}\right)$ are the $\mathrm{L_1}$- and $\mathrm{L_2}$-norm leading to Lasso and Tikhonov regularization respectively.
\section{Proposed Method} 
\label{sec:DL}

From a mathematical point of view, equation~\ref{eq:Tik_ML} can be considered as a function~$f\left(\underline{\mathbf{S}} \right)$, which assigns the signal vector $\underline{\mathbf{S}}$ to the distribution $\underline{\mathbf{F}}$. The same holds true for neural networks, which in general can be described by the following equations:\supercite{Hastie2009TheLearning, Gareth2021AnLearning}
\begin{equation}
    \label{eq:NetZ}
    \underline{\mathbf{Z}}_{\,k} = \sigma \left(\underline{\mathbf{B}}_{\,k} + \underline{\mathbf{W}}_{\,k} \, \underline{\mathbf{Z}}_{\,k-1} \right),
\end{equation}
\begin{equation}
    \label{eq:NetF}
    \underline{\mathbf{F}} = g \left(\underline{\mathbf{Z}}_{\,n}\right),
\end{equation}
where $\underline{\mathbf{Z}}_{\,k}$ is the activation vector of the $k$-th hidden layer including the activations of all neurons in this layer, $\underline{\mathbf{B}}_{\,k}$ is the bias vector of the $k$-th hidden layer consisting of the biases of all neurons in this layer and $\underline{\mathbf{W}}_{\,k}$ the weight matrix of the $k$-th hidden layer containing the weights of all possible connections between the neurons of the $k - 1$-th and $k$-th layer. In this case, the $i$-th row in the matrix $\underline{\mathbf{W}}_{\,k}$ includes the weights of all connections between the $i$-th neuron in the $k$-th layer and all neurons in the $k - 1$-th layer. Furthermore, it holds true that $\underline{\mathbf{Z}}_{\,0} = \underline{\mathbf{S}}$ and $\underline{\mathbf{Z}}_{\,n}$ represents the activations of the last hidden layer $n$. The activation function $\sigma$ is usually chosen to be the rectified linear unit (ReLU), which is given by the following equation:\supercite{GlorotDeepNetworks}
\begin{equation}
    \label{eq:ReLU}
    \sigma \left(x \right) = 
    \begin{cases}
    0, & x < 0 \\
    x, & x \geq 0 \\
    \end{cases}
\end{equation}
For a regression network, the output function $g$ is usually the identity function but if there is a non-negativity constraint present \hl{as is} the case for the reconstruction of distributions the ReLU function is employed instead. The concept of deep learning is then to find the set of weights which minimizes the least-square error between the real distribution and the reconstruction derived from equation \ref{eq:NetZ} and \ref{eq:NetF}. This is usually achieved in an iterative process called network training. After initialisation of the weights which are initially chosen from a normal distribution,\supercite{Glorot2010UnderstandingNetworks, Sutskever2013OnLearning, Mishkin2016AllInit} the mean of the least-square cost function of the training data set is calculated. Afterwards, the gradient with respect to the weights is calculated and the weights are finally updated in negative gradient direction. This iteration cycle is repeated until a convergence or stopping criterion is met. The whole procedure is called gradient descent and is considered as the standard method for network training.\supercite{Hastie2009TheLearning, Gareth2021AnLearning} In recent years, a significant number of improvements to the gradient descent algorithm has been made leading to even more efficient algorithms such as Adam or RMSProp.\supercite{Kingma2014Adam:Optimization, Choromanska2015TheNetworks, Hinton2012ImprovingDetectors} After training, the network can be considered as a non-linear model mapping an input signal $\underline{\mathbf{S}}$ to a distribution $\underline{\mathbf{F}}$ and consequently, can be used to predict a reconstruction of $\underline{\mathbf{F}}$ based on the signal $\underline{\mathbf{S}}$. It is further important to consider here, that the predictions of the model can only be trusted if the signal stems from the same set of distributions as the training data.\supercite{Hastie2009TheLearning} However, if this is ensured and the network is trained sufficiently, it is expected that deep learning should at least show similar performance or even better than regularization techniques such as MTGV and Tikhonov.\supercite{Gareth2021AnLearning} Another factor, which can influence model performance significantly is the choice of network architecture. To identify the best network architecture is not a \hl{straightforward} task and it can be considered more as an engineering problem, which improves due to iterative testing and fine tuning over time instead of being a scientific question with a clear answer to it. A network architecture which is widely used for image regression problems are convolutional neural networks~(CNN). \supercite{Krizhevsky2017ImageNetNetworks, Christian2015GoingConvolutions, He2015DeepRecognition} Those networks are based on the principle that convolutional filters can be used to identify distinct features from the input, which are subsequently forward propagated through the network and finally used to generate an output vector. Hence, during network training the weights of the convolutional filters are estimated, which eventually correlate certain input characteristics with certain output features.\supercite{Gareth2021AnLearning} A particular CNN architecture which was very successfully applied to image-to-image regression problems such as image denoising\supercite{Xie2012ImageNetworks} or image segmentation\supercite{Ronneberger2015U-Net:Segmentation, Long2014FullySegmentation} is Unet. Besides its good performance, another advantage of Unet is that common high-level programming languages such as Matlab, Python or R include default functions to set up the Unet architecture, which allows to obtain a ready-to-train network with a couple of lines of code. Furthermore, the regression problem considered in this paper, which aims to reconstruct a distribution from a signal matrix can be treated as an image-to-image regression problem. The signal $\underline{\mathbf{S}}$ can be easily re-normalised to a grayscale image with maximum signal intensity relating to a white pixel and zero signal to a black one. An analogous transformation holds true for the distribution $\underline{\mathbf{F}}$ and consequently, the reconstruction problem in this paper can be tackled with the Unet architecture in a \hl{straightforward} manner. 
\section{Simulations} 
\label{sec:Sim}

\begin{figure}[t]
    \centering
    \begin{subfigure}[b]{0.24\textwidth}
        \centering
        \includegraphics[keepaspectratio, width=\textwidth]{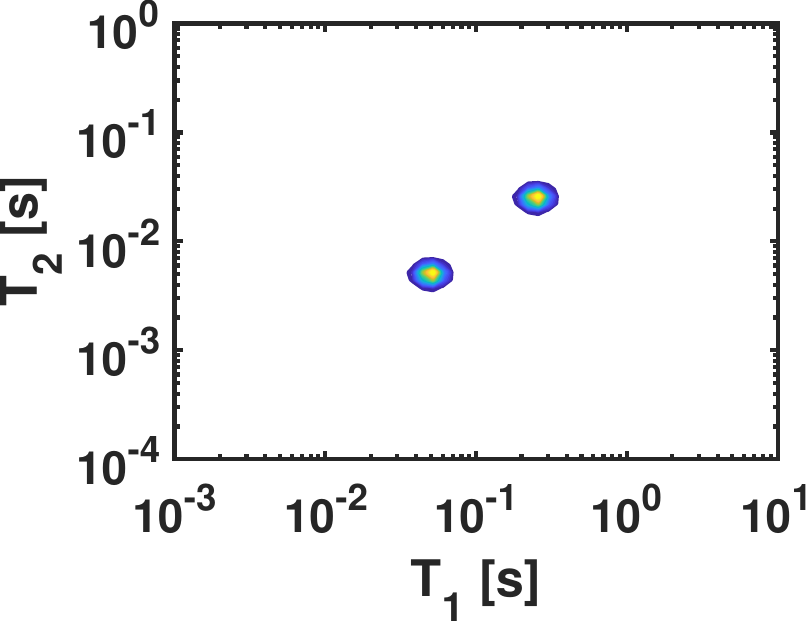}
         \caption{}
    \end{subfigure}
    \begin{subfigure}[b]{0.24\textwidth}
        \centering
        \includegraphics[keepaspectratio, width=\textwidth]{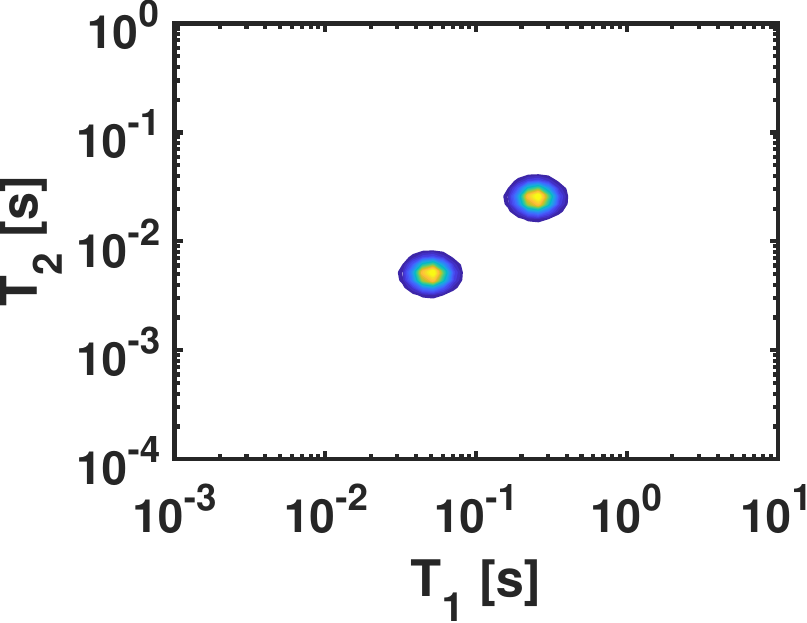}
        \caption{}
    \end{subfigure}
    \begin{subfigure}[b]{0.24\textwidth}
        \centering
        \includegraphics[keepaspectratio, width=\textwidth]{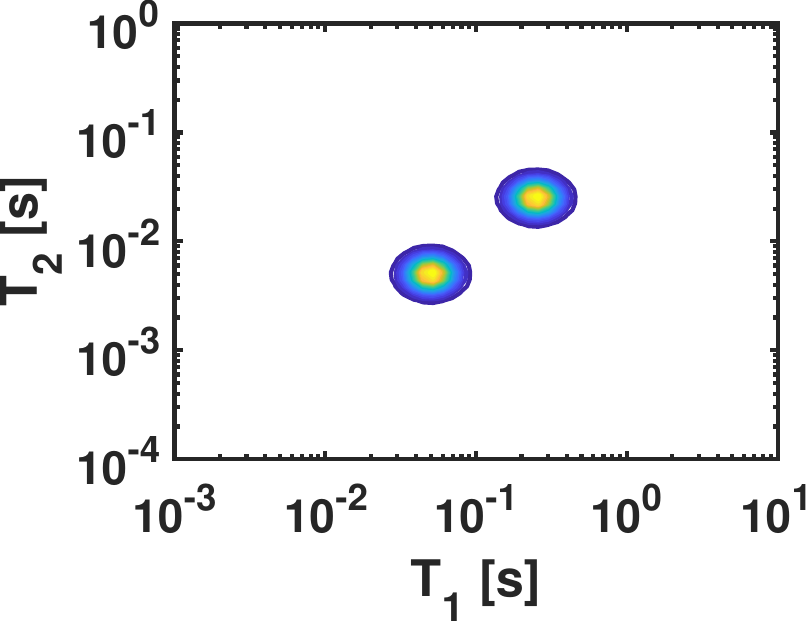}
        \caption{}
    \end{subfigure}
    \begin{subfigure}[b]{0.24\textwidth}
        \centering
        \includegraphics[keepaspectratio, width=\textwidth]{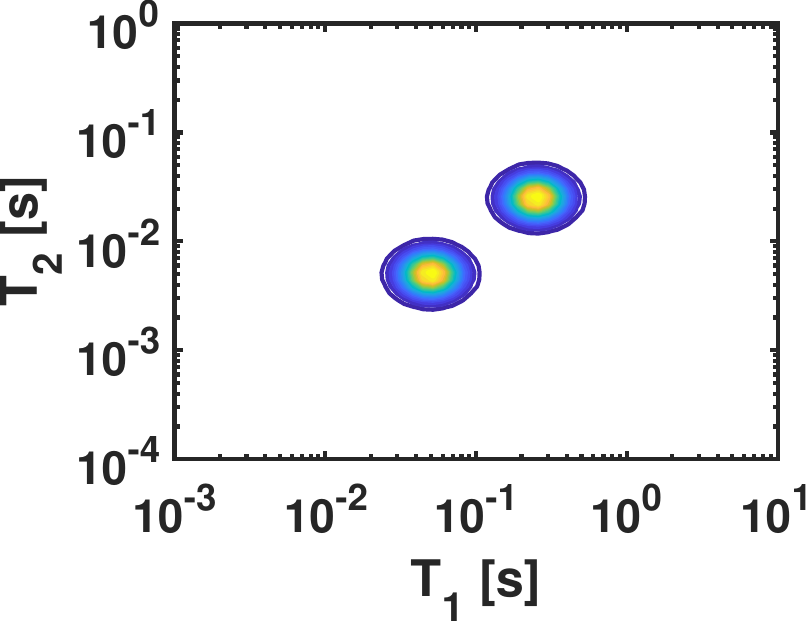}
        \caption{}
    \end{subfigure}
    \\
    \begin{subfigure}[b]{0.24\textwidth}
        \centering
        \includegraphics[keepaspectratio, width=\textwidth]{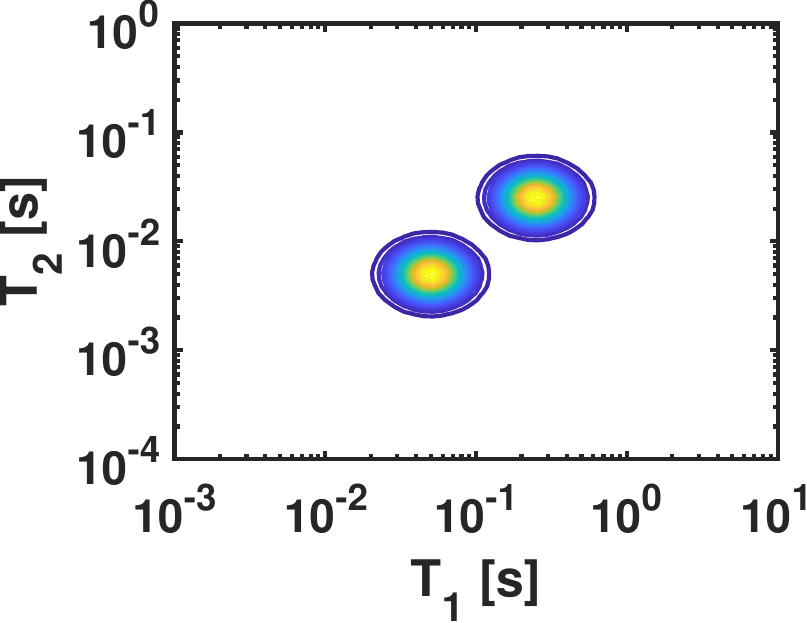}
        \caption{}
    \end{subfigure}
    \begin{subfigure}[b]{0.24\textwidth}
        \centering
        \includegraphics[keepaspectratio, width=\textwidth]{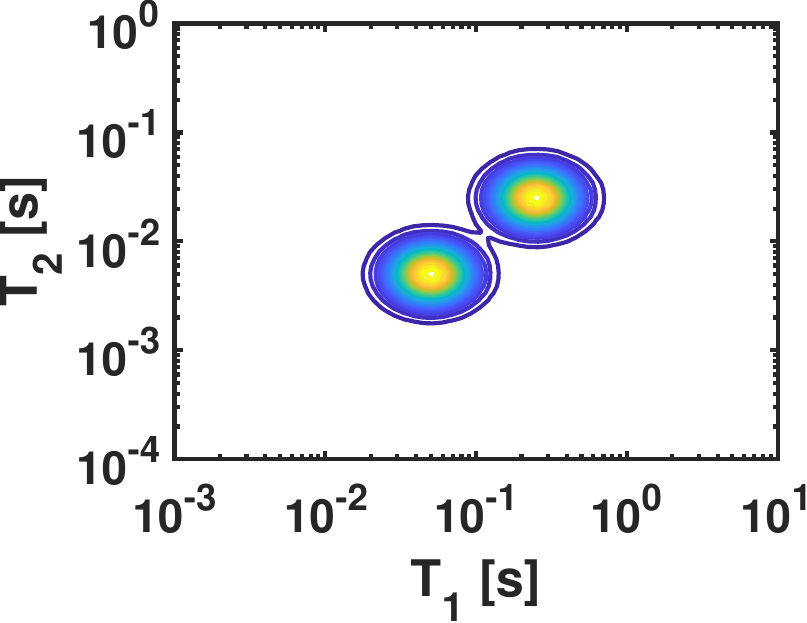}
        \caption{}
    \end{subfigure}
    \begin{subfigure}[b]{0.24\textwidth}
        \centering
        \includegraphics[keepaspectratio, width=\textwidth]{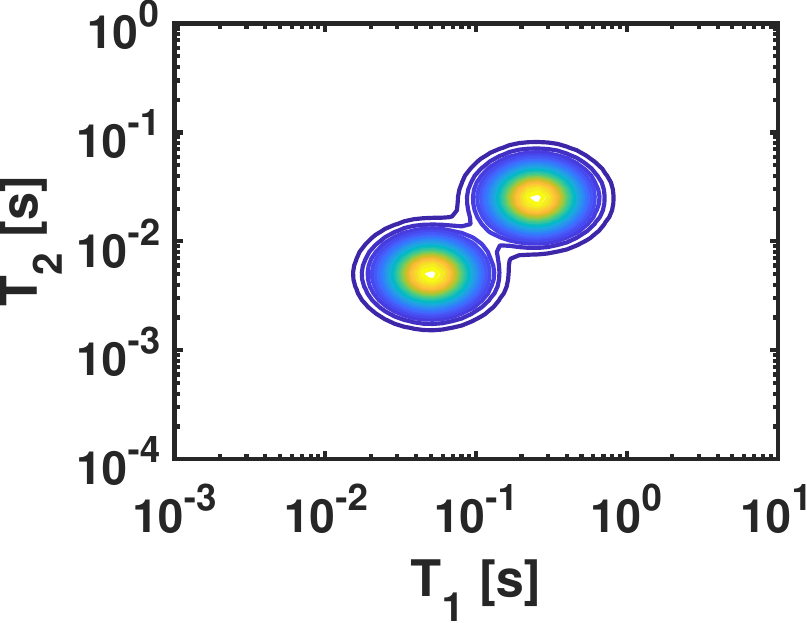}
        \caption{}
    \end{subfigure}
    \begin{subfigure}[b]{0.24\textwidth}
        \centering
        \includegraphics[keepaspectratio, width=\textwidth]{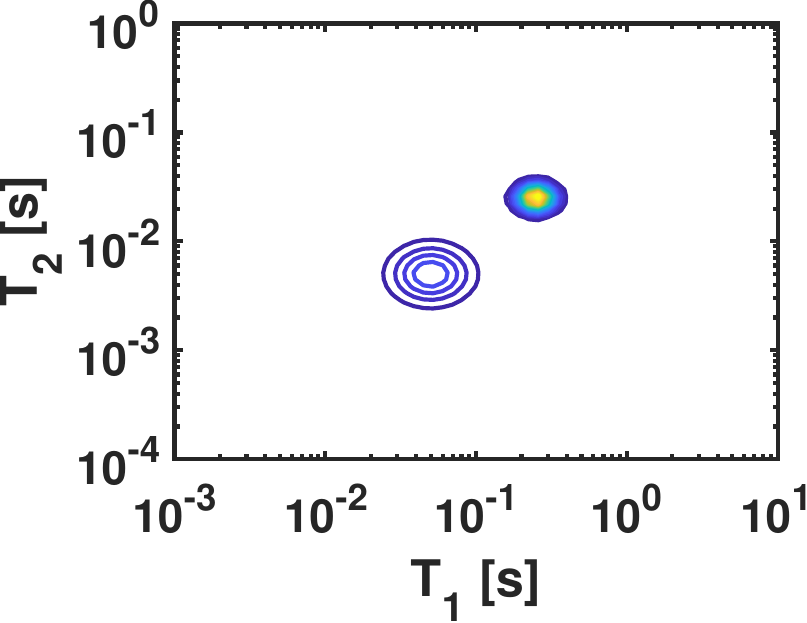}
        \caption{}
    \end{subfigure}
    \caption{Calculated $T_1$-$T_2$-distributions used as input for the simulations of NMR signals employed in this paper. All peaks follow a lognormal distribution and the details can be found in section \ref{sec:Sim}. The labels (a) to (h) correspond to A to H accordingly.}
    \label{fig:Dis} 
\end{figure}

In order to evaluate the performance of the proposed reconstruction method eight $T_1$-$T_2$-distributions have been calculated. All distributions consist of 64 logarithmically spaced data points in both \hl{dimensions} reaching from $10^{-3}$ to $10^{1} \, \mathbf{s}$ in the $T_1$-dimension and from $10^{-4}$ to $10^{0} \, \mathbf{s}$ in the $T_2$-dimension. All distributions include two peaks centred at $T_1 = 0.05 \, \mathbf{s}$, $T_2 = 0.005 \, \mathbf{s}$ and $T_1 = 0.25 \, \mathbf{s}$, $T_2 = 0.025 \, \mathbf{s}$ and are shown in figure~\ref{fig:Dis}. The smoothness of both peaks in the distributions increases from A to G, whereas distribution H includes a sparse as well as a smooth peak. The logarithmic standard deviations associated with the distributions A to H are given in table~\ref{tab:log_std}. The distributions were then used as an input to equation~\ref{eq:SigDis} and random gaussian noise was added to achieve signal-to-noise ratios~(SNR) of 20, 200 and 2000. The resulting signals spanned 64 logarithmically spaced points in both directions reaching from $10^{-3}$ to $10^{1} \, \mathbf{s}$ in the $t_1$-dimension and from $10^{-4}$ to $10^{0} \, \mathbf{s}$ in the $t_2$-dimension
The signals were then fed to the neural network and the distribution reconstructed. For the purpose of performance evaluation, MTGV as well as Tikhonov regularization were also employed to reconstruct the distributions. The hyperparameters for MTGV were estimated through a combination of generalized cross-validation and the Butler-Reeds-Dawson method as it was shown in an earlier publication.\supercite{Beckmann_MTGV} In the case of Tikhonov regularization generalized cross-validation analogously to the MTGV hyperparameter search has been employed.\supercite{Mitchell2012NumericalDimensions} From the $T_1$-$T_2$-distributions $T_1$- and $T_2$-distributions were generated through projection in both dimensions. The reconstructed one- and two-dimensional distributions were then compared between the different reconstruction methods based on a weighted least-squares metric. The objective function $\chi$ defined by Reci \textit{et al.} was used to determine the quality of reconstruction:
\begin{equation} \label{eq:Score}
    \chi = \sum_i \frac{\left(f_{i, true} - f_{i, rec} \right)^2}{\max \left(10^{-4}, f_{i, true} \right)},
\end{equation}
where $f_{i, true}$ is the $i$-th element of the real distribution vector and $f_{i, rec}$ is the $i$-th element of the reconstructed distribution vector. $\chi$ is to be considered as a measure of performance with the best method being the one with smallest value of $\chi$. $\chi$~can then be calculated for every reconstructed distribution and accordingly, the best method identified.
\section{Network Architecture and Training}

To set up the Unet architecture used in this paper, the Matlab function unetLayers with an encoder depth of four was used. From the resulting network, the softmax and the segmentation layer were removed and an additional ReLU and a regression layer were added. 
\begin{table}[t]
\centering
\caption{Lognormal parameters of the employed $T_1\mbox{-}T_2$ distributions.}
\label{tab:log_std}
\resizebox{\textwidth}{!}{
\begin{tabular}{ccccc}
\toprule
Distribution & $T_1$ log mean & $T_1$ log standard deviation & $T_2$ log mean & $T_2$ log standard deviation \\ \midrule
A & -1.30; -0.60 & 0.05; 0.05 & -2.30; -1.60 & 0.05; 0.05 \\
B & -1.30; -0.60 & 0.08; 0.08 & -2.30; -1.60 & 0.08; 0.08 \\
C & -1.30; -0.60 & 0.10; 0.10 & -2.30; -1.60 & 0.10; 0.10 \\
D & -1.30; -0.60 & 0.13; 0.13 & -2.30; -1.60 & 0.13; 0.13 \\
E & -1.30; -0.60 & 0.15; 0.15 & -2.30; -1.60 & 0.15; 0.15 \\
F & -1.30; -0.60 & 0.18; 0.18 & -2.30; -1.60 & 0.18; 0.18 \\
G & -1.30; -0.60 & 0.20; 0.20 & -2.30; -1.60 & 0.20; 0.20 \\ 
H & -1.30; -0.60 & 0.18; 0.08 & -2.30; -1.60 & 0.18; 0.08 \\ \bottomrule
\end{tabular}
}
\end{table}
To generate sufficient training data equation~\ref{eq:SigDis} in combination with a randomised set of $T_1$-$T_2$-distributions was employed. The same number and spacing of data points and identical lower and upper bounds as described in section \ref{sec:Sim} were used and random gaussian noise was added to achieve signal-to-noise ratios between $10$ and $10^4$. The $T_1$-$T_2$-distributions were generated as weighted superpositions of lognormal distributions with up to five components and a logarithmic standard deviation reaching from 0.025 up to 0.25. Based on those parameters $2^{21}$ NMR signals were simulated of which $95 \, \%$ were used for training, $4 \, \%$ for validation and the remaining $1 \, \%$ for testing. The Adam algorithm including $\mathrm{L_2}$-regularization was chosen to minimize the cost function of the neural network with hyperparameters set to the default values in Matlab. A mini-batch size of 222 and an initial learn rate of $10^{-3}$ was used, which was manually decreased down to $10^{-6}$. The data was shuffled every epoch and training progress was validated every 512 iterations. Overall, the network was trained on a single Nvidia GeForce RTX 3070 GPU for approximately 16 hours until convergence was reached, resulting in the performance presented in this publication.
\section{Results and Discussion} 
\label{sec:Res_ML}

\begin{figure}[h!]
    \centering
    \begin{subfigure}[b]{0.3\textwidth}
        \centering
        \includegraphics[keepaspectratio, width=\textwidth]{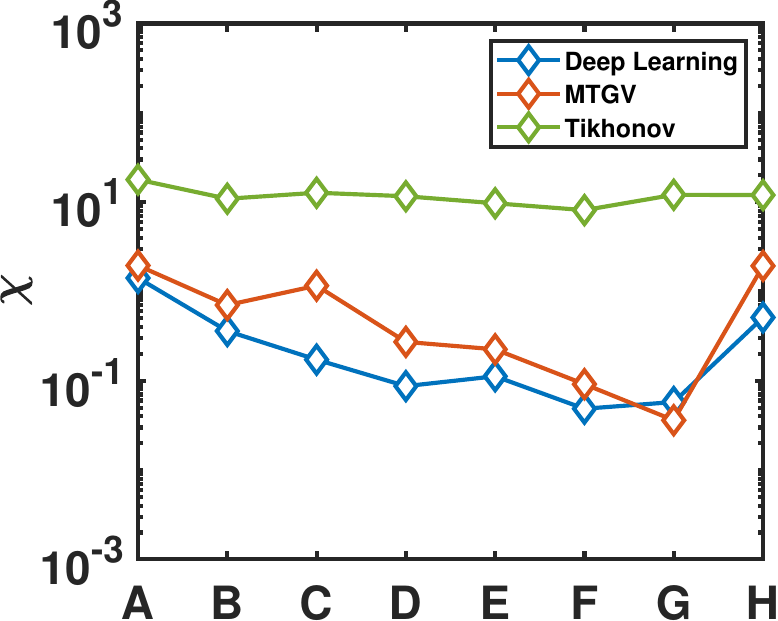}
        \caption{$T_1$-$T_2$, $\mathrm{SNR = 2000}$}
        \vspace{1.1pc}
    \end{subfigure}
    \begin{subfigure}[b]{0.3\textwidth}
        \centering
        \includegraphics[keepaspectratio, width=\textwidth]{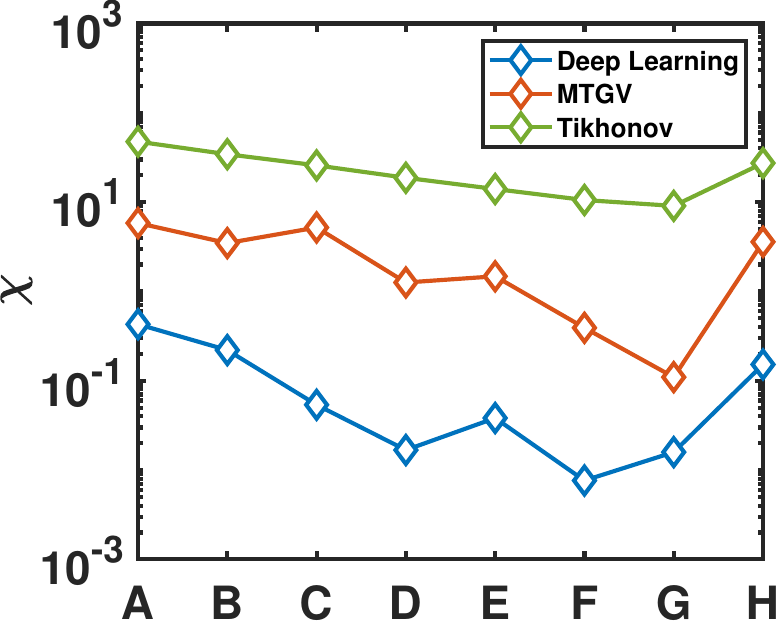}
        \caption{$T_1$, $\mathrm{SNR = 2000}$}
        \vspace{1.1pc}
    \end{subfigure}
    \begin{subfigure}[b]{0.3\textwidth}
        \centering
        \includegraphics[keepaspectratio, width=\textwidth]{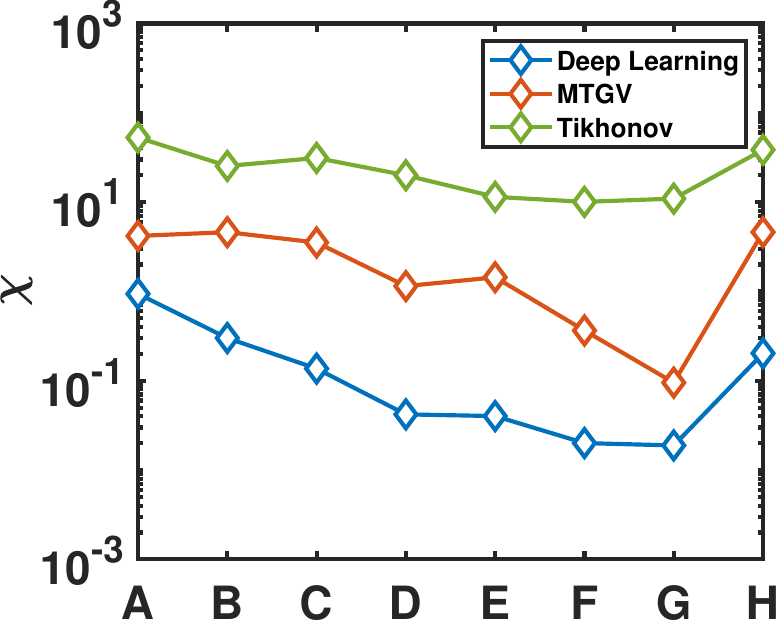}
        \caption{$T_2$, $\mathrm{SNR = 2000}$}
        \vspace{1.1pc}
    \end{subfigure}
    \\
    \begin{subfigure}[b]{0.3\textwidth}
        \centering
        \includegraphics[keepaspectratio, width=\textwidth]{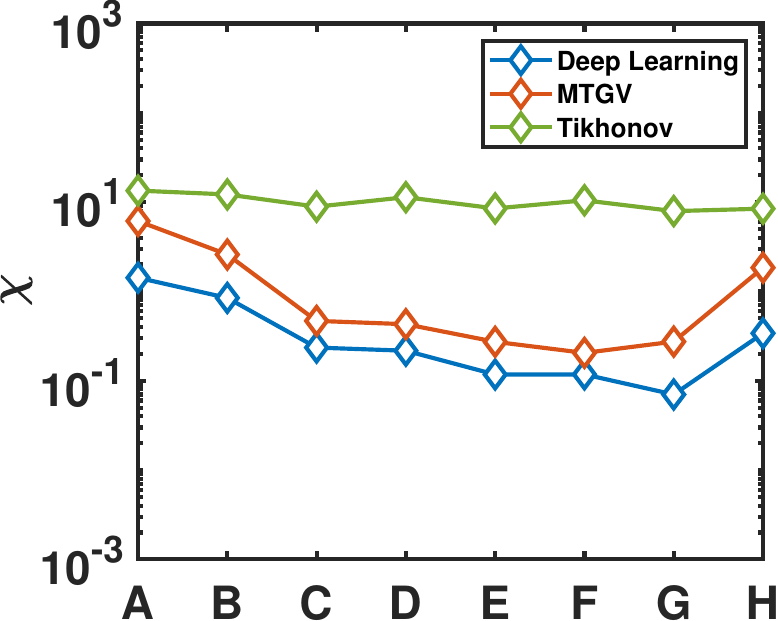}
         \caption{$T_1$-$T_2$, $\mathrm{SNR = 200}$}
         \vspace{1.1pc}
    \end{subfigure}
    \begin{subfigure}[b]{0.3\textwidth}
        \centering
        \includegraphics[keepaspectratio, width=\textwidth]{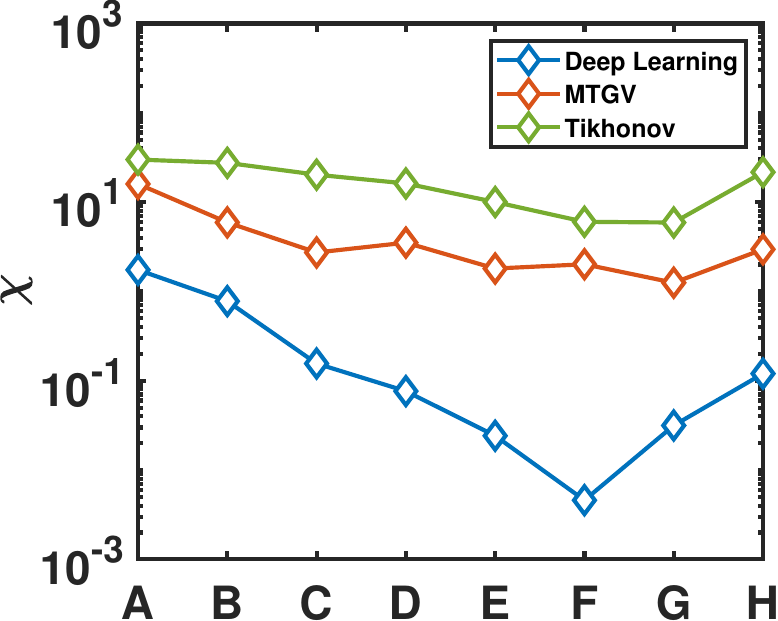}
        \caption{$T_1$, $\mathrm{SNR = 200}$}
        \vspace{1.1pc}
    \end{subfigure}
    \begin{subfigure}[b]{0.3\textwidth}
        \centering
        \includegraphics[keepaspectratio, width=\textwidth]{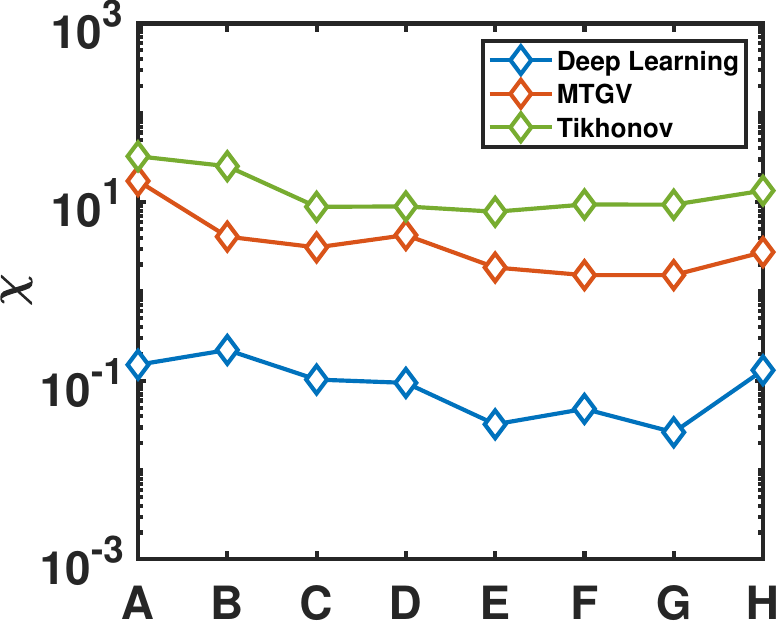}
        \caption{$T_2$, $\mathrm{SNR = 200}$}
        \vspace{1.1pc}
    \end{subfigure}
    \\
    \begin{subfigure}[b]{0.3\textwidth}
        \centering
        \includegraphics[keepaspectratio, width=\textwidth]{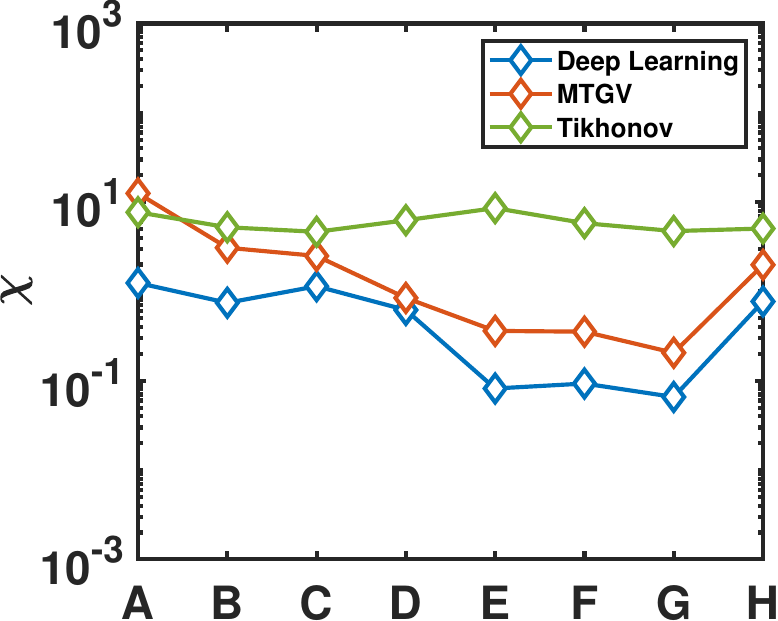}
        \caption{$T_1$-$T_2$, $\mathrm{SNR = 20}$}
    \end{subfigure}
    \begin{subfigure}[b]{0.3\textwidth}
        \centering
        \includegraphics[keepaspectratio, width=\textwidth]{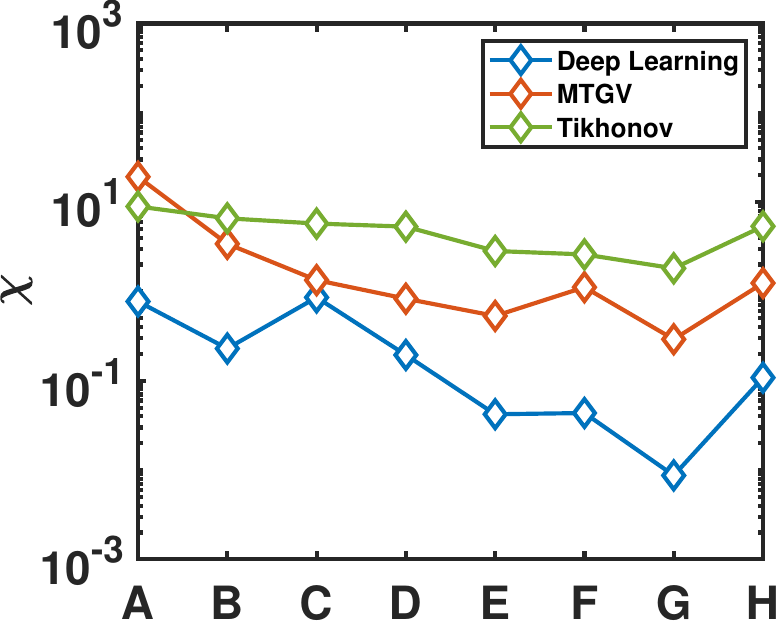}
         \caption{$T_1$, $\mathrm{SNR = 20}$}
    \end{subfigure}
    \begin{subfigure}[b]{0.3\textwidth}
        \centering
        \includegraphics[keepaspectratio, width=\textwidth]{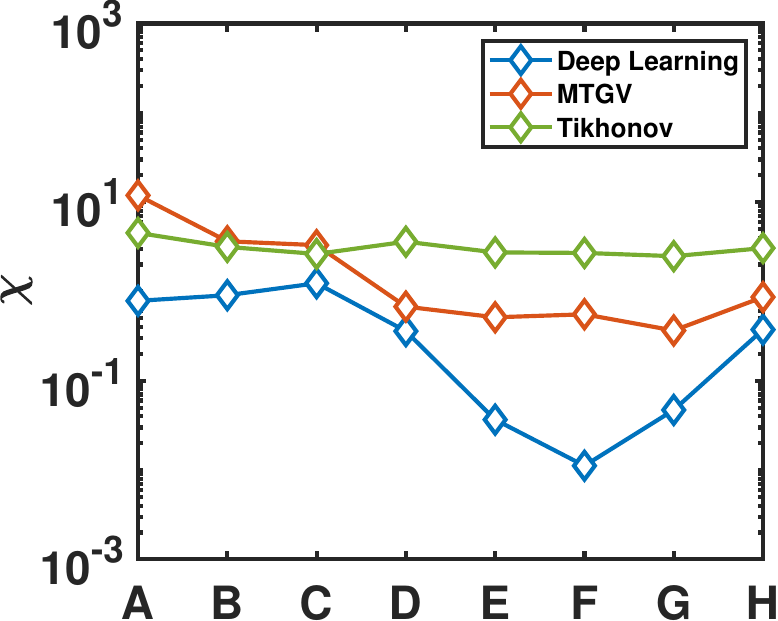}
         \caption{$T_2$, $\mathrm{SNR = 20}$}
    \end{subfigure}
    \caption{$\chi$-scores obtained for the reconstructed distributions generated via the deep learning approach presented in this paper.}
    \label{fig:2000} 
\end{figure}

$\chi$ was calculated for the distributions A to H including all three reconstruction methods as well as all three signal-to-noise ratios considered in this publication. The results are shown in figures~\ref{fig:2000}. Comparing the $\chi$-scores of all three reconstruction methods, it becomes evident that deep learning outperforms MTGV and Tikhonov regularization with only one exception. In the two-dimensional case of the distribution G, MTGV regularization shows slightly better results than deep learning but only if the signal-to-noise ratio is 2000. In all other instances, independent whether two- or one-dimensional data sets are used deep learning provides a better reconstruction than both regularization methods. A further comparison between reconstructions differing in signal-to-noise ratio only shows that for all tested methods the quality of reconstruction is only weakly influenced by a low signal-to-noise ratio. Overall, only the signals leading to distributions C and D with a signal-to-noise ratio of 20 achieve a considerably higher $\chi$-score than their high signal-to-noise counterparts. Comparing the two-dimensional reconstructions with the one-dimensional projections, it becomes evident, that in the two-dimensional case overall MTGV regularization performs only slightly worse than deep learning but clearly better than Tikhonov. In contrast, the reconstructions of the one-dimensional distributions show a very distinct overall performance advantage in favour of deep learning, whereas especially with increasing noise the performance of MTGV and Tikhonov regularization becomes more similar. Another important factor to judge the performance of an inversion method is the running time of the algorithm. Passing a signal vector through a pre-trained network is usually fast and consequently, on a Windows 10 desktop computer equipped with 32~GB of RAM and an AMD Ryzen~7 5800X processor, the reconstruction of the signals described in section \ref{sec:Sim} took less than one second, which is several orders of magnitude faster than reconstruction via MTGV or Tikhonov regularization.  
To be more accurate, reconstruction of the same signals took in the case of MTGV \hl{on} average about twelve minutes including a full hyperparameter search, whereas Tikhonov achieves a running time roughly one order of magnitude smaller than MTGV. This is still approximately two orders of magnitude longer than the neural net highlighting the great efficiency of the deep learning approach regarding running time. From a perspective of routine application, inversion via deep learning has the further advantage that with a pre-trained network no hyperparameters or convergence criteria have to be estimated. This means that the deep learning based approach can be easily used by application focused scientists without profound knowledge of the inversion algorithm, whereas for MTGV and Tikhonov regularization some experience regarding the selection of hyperparameters and convergence criteria is usually necessary. A further benefit is that the proposed inversion method is invariant regarding translation. This means the network trained for this publication can also be used for inversion of signals from other types of experiments such as $T_1$-$D$-correlation measurements. The only prerequisite for this is that the signal vector has the same length and provides data points with an identical relative logarithmic distance between two adjacent vector entries as the training data sets employed in this paper. From this, one downside of inversion via deep learning can be inferred directly, which is a fixed length of the input and output vector. A further disadvantage is that the predictions of a neural net can only be expected to be reliable if the to be inverted signal stems from similar distributions as the signals included in the training data set. However, both of those issues can be easily overcome by re-training of a pre-trained network or data augmentation. For instance, missing data points in a signal could be interpolated from spline fitting of the initial vector, network layers can be added or removed to fit the needed input or output size and the network can be re-trained on a data set appropriate for the inversion problem of interest. In this case, re-training of a network is expected to be fast because at least a part of the weights of the network employed in this publication could be re-used, which increases convergence speed in a quick and straightforward manner. In addition, further performance improvements seem to be possible, but this is expected to require changes to the network architecture, which are \hl{beyond} the scope of this paper.
\section{Conclusion}

In this paper, deep learning was successfully employed as a method for inversion of NMR signals. The proposed inversion network makes use of the Unet architecture, which is routinely applied for conceptually similar regression problems such as image denoising and image segmentation. The inversion performance was tested on simulated NMR signals with signal-to-noise ratios of 20, 200 and 2000 stemming from eight superpositions of lognormal distributions differing in sparsity and smoothness. Based on the weighted least-squares metric proposed in Reci's original MTGV work, the results were compared with reconstructions derived from MTGV as well as Tikhonov regularization. This comparison showed that inversion via deep learning outperformed both regularization techniques in nearly all instances. It was further shown that inversion via deep learning is at least two orders of magnitude faster than MTGV- and Tikhonov regularization. Finally, the benefits of deep learning inversion regarding routine usability were highlighted and possible shortcomings of the method were discussed.

\printbibliography[title=References]

\end{document}